\documentclass[a4paper,twoside]{article}
\newcommand{\affil}[1]{$^{\rm #1}$}
\usepackage[authoryear]{natbib}
\bibpunct{(}{)}{;}{a}{}{,}
\usepackage{graphicx}
\date{} 
\title{\large\bf\flushleft Detailed Chromospheric Activity Nature of KIC 9641031}
\author{\parbox{\textwidth}{\flushleft
\vspace{-0.5cm}
{\it Ezgi YOLDA\c{S}\affil{1}, Hasan Ali DAL\affil{1},\affil{2}}\\
\vspace{0.4cm}
{\small \affil{1}\,Department of Astronomy and Space Sciences, University of Ege, Bornova, 35100 ~\.{I}zmir, Turkey}\\
{\small \affil{2}\,Corresponding Author, Email: ali.dal@ege.edu.tr}}}
\begin{document}
\twocolumn[
\begin{changemargin}{.8cm}{.5cm}
\begin{minipage}{.9\textwidth}
\vspace{-1cm}
\maketitle
%
%
\small{\bf Abstract:} This study depends on KIC 9641031 eclipsing binary system with a chromospherically active component. There are three type variations, such as geometrical variations due to eclipses, sinusoidal variations due to the rotational modulations and also flares, in the light curves obtained with the data taken from the Kepler Mission Database. Taking into account results obtained from KIC 9641031's observations in the Kepler Mission Database, we present and discuss the details of chromospheric activity. The sinusoidal light variations due to rotational modulation and the flare events were modelled separately. 92 different data subsets separated using the analytic models described in the literature were modelled separately to obtain the cool spot configuration. It is seen that just one component of the system is chromospherically active star. On this component, there are two active regions separated by about $180^\circ$ longitudinally between the latitudes of $+50^\circ$ and $+100^\circ$, whose locations and forms are rapidly changing in short time intervals. 240 flares were detected and their parameters were computed. Using these parameters, the One Phase Exponential Association (hereafter OPEA) function model was derived, in which the $Plateau$ value as a saturation level of the flare-equivalent duration was found to be 1.232$\pm$0.069 s for KIC 9641031, and $half-life$ parameter was found to be 2291.7 s. The flare frequency $N_{1}$ was found to be 0.41632 $h^{-1}$, while the flare frequency $N_{2}$ was found to be 0.00027. Considering these parameters together with the orbital period variations demonstrates that the period variations directly depend on chromospheric activity. Comparing the system with its analogue it is seen that the activity level of KIC 9641031 is remarkablely lower than the others. Although activity level of the system is very low, but the sensitivity of the Kepler Satellite is so high that the activity on the component gets into clearly observable range.\\


\medskip
\medskip
\end{minipage}
\end{changemargin}
]
\small

\section{Introduction}

The flare stars as known UV Ceti are generally young dwarf stars from the spectral types G, K and M with emission lines in their spectra, such as dMe, which are just coming to the main sequence. As it was indicated by \citet{Mir90} and \citet{Pig90}, the population rate of the dwarf stars having flare phenomenon is generally high in both the open clusters and the associations. As it is expected due to the Skumanich's law, the population rate of the stars showing flare activity has a reducing trend, while the age of the cluster gets older \citep{Sku72, Pet91, Sta91, Mar92}.

Strengthening stellar chromospheric activity with increasing the rotation velocity causes much more mass loss from the star. It is known that the stellar mass loss rate of V Ceti type stars is about $10^{-10}$ $M_{\odot}$ per year due to flare like events, while the solar mass loss rate is about $2\times10^{-14}$ $M_{\odot}$ per year \citep{Ger05}. In the case of UV Ceti type flare stars, this high mass loss rate clarifies how they lose large part of their angular momentum in their main sequence stages \citep{Mar92}. However, the flare activity mechanism resulting high level mass loss has not been completely explained by any theory yet.

Flare events observed on UV Ceti type stars are generally explained with the classical theory of the solar flare. The highest energy detected from the most powerful flares, known as two-ribbon flares, occurring on the sun is found to be $10^{30}$ - $10^{31}$ erg \citep{Ger05, Ben08}. This level is also observed for RS CVn type active binaries \citep{Hai91}. However, in the case of dMe stars, this flare energy level varies from $10^{28}$ erg to $10^{34}$ erg \citep{Ger05}. Moreover, some stars of the young clusters such as the Pleiades cluster and the Orion association exhibit some powerful flare events, which energies reach $10^{36}$ erg \citep{Ger83}. In brief, the stars from different type exhibit some flares with different energy levels. Comparing the sun with a dMe stars, it is seen that there are some remarkable difference between both their flare energy levels and mass loss rate per year. Nevertheless, the flare events occurring on a dMe star are generally tried to explain by the classical theory of solar flare. Therefore, the primary energy source in the flare events is magnetic reconnection processes in the principle \citep{Hud97, Ger05}. To reach the real situation, examining the flare events occurring on the different type stars, all the differences and similarities should be demonstrated. Then, it should be identified which parameters such as singularity, binarity, mass, age, ect., cause these differences and similarities.

In this study, both flare and stellar cool spot activities observed from a well known eclipsing binary star FL Lyr as a different system apart from classical UV Ceti type flare stars from the spectral type dMe were analysed and modelled. Then, the results were compared with the chromospheric activity behaviour observed on its analogue. The photometric data used in the analyses and models were taken from the Kepler Mission Database \citep{Sla11, Mat12}. In the database, FL Lyr is listed with its catalog number as KIC 9641031.

KIC 9641031 has been listed in the Youngest-Field Star Catalog by \citet{Gui09}, though the system is not young. \citet{Bro10} given the age of the system between 3.05 Gy and 15.25 Gy. \citet{Mor35} listed KIC 9641031 as a variable star for the first time in the literature, while \citet{Str50} classified the system as a spectroscopic binary with a single line. \citet{Mac52} and \citet{Min66} observed the system photometrically, especially in the narrow-bands.

Although the spectral type of the system is given as $F8V+G8V$ in some catalogue \citep{Eke14}, the temperatures of the components are given in a large range in the general literature. The temperatures of the primary component are varying from 5724 K \citep{Gui09} to 6412 K \citep{Eke14} in the literature. Its temperature is generally accepted to be 6150 K \citep{Bro10, Eke14}. Similarly, the temperatures stated for the secondary component is changing from 5080 K \citep{Gui09} to 5506 K \citep{Pop86}. It is accepted as 5300 K in general \citep{Bro10, Arm14}. The first light curve analysis of the system was made by \citet{Jur76}. In the study, some parameters such as the fractional radii ($r$), the inclination ($i$) of the system, the fractional luminosity ($L_{1}$) of the primary component were computed. The light curves were analysed together with the radial velocity curve simultaneously by \citet{Lac85} for the first time, thus the first approximations were obtained for the masses and radii of the components. The component masses were given as $M_{1} =$ 1.218 $M_{\odot}$ and $M_{2} =$ 0.958 $M_{\odot}$, the radii of the components were given as $R_{1} =$ 1.283 $R_{\odot}$ and $R_{2} =$ 0.963 $R_{\odot}$ by \citet{Eke14}. The semi-major axis ($a$) of the system was given as $a = 9.17$ $R_{\odot}$ and the inclination ($i$) of the system was given as $86.3^\circ$ in the same catalogue. According to this view, the components are comparatively far away from each other. Analysing the spectroscopic observations of the system, \citet{Cri65} asserted existence of some clues for the third body, but \citet{Lac79} did not find any sign for the third body though they obtained the double-lined radial velocity curves of the system. \citet{Pop86} also obtained the double-lined radial velocity curves of the binary.

As it is seen from the literature, KIC 9641031 is a well studied eclipsing binary for some decades, but the chromospheric activity nature of the system has not been exhaustively studied yet. \citet{Bot78} stated that one of the components exhibits magnetic activity. Then, \citet{Balo15} listed some statistical information about flare activity occurring on the components some decades later. However, there is no detailed study about its chromospheric activity nature or any comparison with the other chromospherically active stars about the similarities or differences between them.

In this study based on the Kepler Mission short-cadence light curves, to understand the magnetic activity nature of the system, we analysed and modelled KIC 9641031's light variations due to the stellar cool spots and the flare events occurring on the components as the chromospheric activity indicators. Then, comparing with the results obtained from other stars, our results were discussed for both the cool spots and the flares.

\section{Data and Analyses}

More than 150.000 targets have been observed in the Kepler Mission, which is aimed to find out exo-planet \citep{Bor10, Koc10, Cal10}. The observations in this Mission have the highest quality and sensitivity ever reached in the photometry \citep{Jen10a, Jen10b}. In this case, lots of variable targets such as new eclipsing binaries, etc. have been also discovered apart from the exo-planets \citep{Sla11, Mat12}. Many targets among these newly discoveries are single or double stars exhibiting chromospheric activity. Some of the double stars are the eclipsing binaries \citep{Balo15}.

The photometric data of KIC 9641031 were taken from the Kepler Mission Database. All the light curves obtained from these data are together with error bars shown in Figure 1. The error of each point is 0.00008 mag or smaller. Instead of long-cadence light curves, the short-cadence data were used in the analyses and models \citep{Sla11, Mat12}. In order to analysis the cool spots, the flare events and the orbital period variation (O-C), the data were arranged as the suitable formats.

\subsection{Flare Activity and the OPEA Model}

To determine the light variations just due to the flare events, in the first step, all the light variations except the flares were removed from the general light variation. For this purpose, the data of all the primary minima observations between the phases of 0.96-0.04 and all the secondary minima observations between 0.46 and 0.54 in phase were removed from the general light curve data. By the way, all the deviated observations with large error due to the technical problems were also removed from the data sets. And thus, the pre-whitened light curves were obtained.

In the second step, to determine the basic flare parameters such as the first point of the flare beginning, the last point of the flare end and the flare energy, the quiescent levels for each flare should be derived from the residual data of the pre-whitened light curves. However, it was seen that there is a sinusoidal variation due to rotational modulation variation because of the cool spots occurring on the components. Therefore, considering the pre-whitened light curve just out-of-flare, the light variations seen due to rotational modulation were modelled with the Fourier transform, using the least-squares method. Thus, these synthetic models lead us to definite the quiescent levels for each flare at the same time of that flare. Using these synthetic models as the quiescent levels, the parameters of the flares were computed. Two flare light curves taken from the observation data and the quiescent levels derived for these flares are shown in Figures 2 and 3.

Using the synthetic models derived for the quiescent levels, flare rise times ($T_{r}$), decay times ($T_{d}$), amplitudes of flare maxima, flare equivalent durations ($P$) were computed, after defining both the flare beginning and the end for each flare. In total, 240 flares were detected from the available data in the Kepler Mission Database. All the computed parameters are listed in Table 1 for these 240 flares. In the table, flare maximum times, equivalent durations, rise times, decay times and amplitudes of flare maxima are listed from the first column to the last, respectively.

The equivalent durations of the flares were computed using Equation (1) taken from \citet{Ger72}:
\begin{center}
\begin{equation}
P = \int[(I_{flare}-I_{0})/I_{0}] dt
\end{equation}
\end{center}
where $I_{0}$ is the flux of the star in the observing band while in the quiet state. $I_{0}$ was computed using the models derived with the Fourier transform. $I_{flare}$ is the intensity observed at the moment of flare. Finally, $P$ is the flare-equivalent duration in the observing band. In this study, the flare energies were not computed to be used in the following analyses due to the reasons described in detail by \citet{Dal10, Dal11}. Before computing the flare-equivalent duration, comparing synthetic models of the quiescent levels with the observations, the times of flare beginning, end and also maximum were defined, and then computing time differences of these points, the flare rise and decay times were calculated.

Examining the relationships of the flare parameters among each other, it is seen that the distributions of flare equivalent durations on the logarithmic scale versus the flare total durations are varying according to a rule. The distributions of flare equivalent durations on the logarithmic scale can not be higher than a specific value for the star, and it is no matter how long the flare total duration is. Using the SPSS V17.0 \citep{Gre99} and GrahpPad Prism V5.02 \citep{Daw04} programs, \citet{Dal10, Dal11} indicated that the best function is the OPEA Model to fit the distributions of flare equivalent durations on the logarithmic scale versus flare total durations. The OPEA function has a $Plateau$ term, and this makes it a special function in the analyses. The OPEA function is defined by Equation (2):

\begin{center}
\begin{equation}
y~=~y_{0}~+~(Plateau~-~y_{0})~\times~(1~-~e^{-k~\times~x})
\end{equation}
\end{center}
where the parameter $y$ is the flare equivalent duration on a logarithmic scale, the parameter $x$ is the flare total duration, according to the definition of \citet{Dal10}, and the parameter $y_{0}$ is the flare-equivalent duration in on a logarithmic scale for the least total duration. In other words, the parameter $y_{0}$ is the least equivalent duration occurring in a flare for a star. Here is an important point that the parameter $y_{0}$ does not depends on only flare mechanism occurring on the star, but also depends on the sensitivity of the optical system used for the observations. The parameter $Plateau$ value is upper limit for the flare equivalent duration on the logarithmic scale. \citet{Dal11} defined $Plateau$ value as a saturation level for a star in the observing band.

Using the least-squares method, the OPEA model was derived for the distributions of flare equivalent durations on the logarithmic scale versus the flare total durations. The derived model is shown in Figure 4 together with the observed flare equivalent durations, while the parameters computed from the model are listed in Table 2. The $span$ value listed in the table is difference between $Plateau$ and $y_{0}$ values. The $half-life$ value is half of the first $x$ value, at which the model reaches the $Plateau$ value. In other words, it is half of the minimum flare total time, which is enough to the maximum flare energy occurring in the flare mechanism.

It was tested by using three different methods, such as the D'Agostino-Pearson normality test, the Shapiro-Wilk normality test and also the Kolmogorov-Smirnov test, given by \citet{Dag86} to understand whether there are any other functions to fit the distributions of flare equivalent durations on the logarithmic scale versus the flare total durations. In these tests, the probability value called as $p-value$ was found to be $p-value < 0.001$. This means that there is no other function to model the distributions of flare equivalent durations \citep{Mot07, Spa87}.

KIC 9641031 was observed as long as 576.47474 days from JD 2454964.50251 to JD 2456424.01145 without any remarkable interruptions. In total, significant 240 flares were detected in these observations. The total flare equivalent duration computed from all the flares was found to be 556.81321 s (0.15467 hours). \citet{Ish91} described two frequencies for the stellar flare activity. These frequencies are defined as given by Equations (3) and (4):

\begin{center}
\begin{equation}
N_{1}~=~\Sigma n_{f}~/~\Sigma T_{t}
\end{equation}
\end{center}

\begin{center}
\begin{equation}
N_{2}~=~\Sigma P~/~\Sigma T_{t}
\end{equation}
\end{center}
where $\Sigma n_{f}$ is the total flare number detected in the observations, and $\Sigma T_{t}$ is the total observing duration, while $\Sigma P$ is the total equivalent duration obtained from all the flares. In this study, $N_{1}$ frequency was found to be 0.41632 $h^{-1}$, while $N_{2}$ frequency was found to be 0.00027.

\subsection{Rotational Modulation and Stellar Spot Activity}

The light curves of the system indicate the existence of the sinusoidal variations out-of-eclipses. Considering the temperatures of the components, the sinusoidal variations are more likely caused by rotational modulation due to the cool stellar spots. As it is clearly seen from the temperature ranges given in the literature for the components, both the primary and secondary components are potential candidates to exhibit the chromospheric activity. In this study, we assumed that the secondary component is a chromospherically active star. To demonstrate just the sinusoidal variations, both all the minima due to the eclipses of the components and all the flare as seen sudden - rapid increasing in the light were removed from general light curves, thus the remaining light curves were obtained, which is hereafter called as the pre-whitened light curves. Comparing the variations seen in the pre-whitened light curves cycle by cycle according to the orbital period of the system, it was seen that both the phases and levels of maxima and minima are rapidly changing from one cycle to the next. The situation is clearly an indicator for the rapid evolution of the magnetically active regions on the components. In this case, all the pre-whitened light curves can not be modelled as just one data set, because of this, the data set of whole pre-whitened light curves were separated to sub-data sets. In this process, the consecutive cycle data, which have almost the same phase distributions and brightness levels, were arranged as one sub-data set. Thus, all the available data were arranged as 92 sub-data sets and each sub-data set was individually modelled.

To find out the parameters of spot distribution on the stellar surface such as the spot radius, latitude and especially longitude, we modelled the sub-data sets under some assumptions using the SPOTMODEL program \citep{Rib02, Rib03}. In this program, the analytic models of \citet{Bud77} were used to model the sinusoidal variations out-of-eclipses. The program needs two-band observations or spot temperature factor ($kw=[T_{spot}/T_{surface}]^2$) parameter. However, the available data in the Kepler Mission Database contain only monochromatic observations. Therefore, considering the clues of the spot activity for this system stated firstly by \citet{Bot78} and also the results obtained from the light curve analyses of its analogue systems \citep{Cla01, Tho08}, it was assumed that the secondary component exhibits chromospheric activity.

\begin{center}
\begin{equation}
\label{eq:five}
L(\theta)= A_{0} ~ + ~ \sum_{\mbox{\scriptsize\ i=1}}^N ~ A_{i} ~ cos(i \theta) ~ + ~ \sum_{\mbox{\scriptsize\ i=1}}^N ~ B_{i} ~ sin(i \theta)
\end{equation}
\end{center}

Considering the coefficients $A_{i}$ and $B_{i}$, the dominant term is $\cos (i \theta)$ for the first ($i=1$) and second ($i=2$) orders in the analysis with the Fourier transform described by Equation (5). According to \citet{Hal90}, the $\cos (i \theta)$ term is an indicator for the spotted areas on the surface of a star. In this case, the previous analyses with the Fourier transform indicate that there are two active regions on the active component. Considering the results of the analogue systems, it was assumed that the spot temperature factors are in the range of $kw = 0.70 - 0.95$ for these cool spots. The initial models by changing the factors from 0.70 to 0.95 reveal that the best solutions are obtained if the spot temperature factor is taken as $kw = 0.75$ for the primary spot (Spot 1), while it is taken as $kw = 0.85$ for the secondary one (Spot 2). Consequently, it was assumed that the spot temperature factors are constant parameters for each sub-data set, and they are taken as $kw = 0.75$ for the first spot and $kw = 0.85$ for the second spot in the each model. Finally, taking the spot temperature factor as a constant, the longitudes ($I$), latitudes ($b$) and radii of the spots ($g$) parameters are taken as the adjustable free parameters in the each model.

Five examples selected from different time intervals among all 92 models derived by SPOTMODEL program are shown in Figure 5. In the figure, both the model fits and the cool spot distributions on 3D surface are seen side by side for these five selected sub-data sets. All the spot parameters derived by SPOTMODEL program are also listed in Table 3. In the table, the average Heliocentric Julian Date of the time interval for each sub-data set (HJD), spot latitudes ($b$), radii of the spots ($g$) and spot longitudes ($l$) are listed from the first column to the last, respectively.

The variations of spot latitude ($b$), spot radius ($g$) and spot longitude ($l$) values versus the time are shown in Figure 6. In this point, it must be noted that if it was assumed that chromospherically active star is not the secondary component, but the primary one, there would be no distinctive changes in the values of spot latitudes ($b$), radii of the spots ($g$) and spot longitudes ($l$). This is because the surface temperatures of the both components are so close to the each other.

\subsection{Orbital Period Variation}

The minima times were computed with a script depending on the method described by \citet{Kwe56}. Each minimum in the light curves was separately fitted with the high order spline functions. Using these fits, the minima times were computed from the available short cadence detrended data of the system in the Kepler Mission Database \citep{Sla11, Mat12} without any extra correction on these detrended data. For all the minima times, the differences between the observations and the calculations were computed to determine the residuals $(O-C)_{I}$. Some minima times have very large error, for which the minima light curves were examined. It was seen that there is a flare activity during these minima, then, these minima times were removed from analyses. Finally, 532 minima times were obtained from the observations in the Kepler Mission. Using the regression calculations, a linear correction was applied to the differences, and the $(O-C)_{II}$ residuals were obtained. After the linear correction on $(O-C)_{I}$, new ephemerides were calculated as following:

\begin{center}
\begin{equation}
JD~(Hel.)~=~24~54954.13348(4)~+~2^{d}.1781543(1)~\times~E.
\end{equation}
\end{center}

The minima times, epoch, minimum type, $(O-C)_{I}$ and $(O-C)_{II}$ residuals are listed in Table 4, respectively. The error of each minimum time in the table is 0.00001 day or smaller. The $(O-C)_{II}$ residual variations versus time are shown in Figure 7. As seen from the figure, the $(O-C)_{II}$ residuals exhibit some distorted sinusoidal variations in opposite directions relative to each other. A similar phenomenon has been recently demonstrated for chromospherically active other systems by \citet{Tra13} and \citet{Bal15}.

\section{Results and Discussion}

The analyses of data taken from the Kepler Mission Database \citep{Sla11, Mat12} indicated that KIC 9641031 is a chromospherically active system. However, to reach the certain results about the activity level of the system, it needs to compare the results with its analogues. Considering the temperatures of the components, both components seem to be potential candidates to exhibit the chromospheric activity. However, in this study, we assumed that the secondary component is a chromospherically active star. Using the calibrations given by \citet{Tok00}, we derived $B-V$ color index for the secondary component depending on its temperature generally accepted as 5300 K. According to the calibrations, the $B-V$ color index of a main sequence star with the temperature of 5300 K was found to be $0^{m}.74$.

KIC 9641031 was observed as long as 576.47474 hours between JD 2454964.50251 and JD 2456424.01145. We found 240 flares from these observations. Apart from other flare parameters, the flare frequencies were also computed. $N_{1}$ flare frequency was computed as 0.41632 $h^{-1}$, while $N_{2}$ frequency was found to be 0.00027. Comparing these values with the frequencies found for UV Ceti type flare stars in a wide spectral range, from spectral type dK5e to dM6e, it is clearly seen that the flare energies obtained from KIC 9641031 are remarkably lower than its analogues. In addition, the number of the flare occurring on the star per hour is also remarkably less than its analogues. As it can be seen from the literature, for example, the observed flare number per hour for UV Ceti type single stars was found to be $N_{1} =$ 1.331 $h^{-1}$ in the case of AD Leo, while it was found to be $N_{1} =$ 1.056 $h^{-1}$ for EV Lac. Moreover, $N_{2}$ frequency was found to be 0.088 for EQ Peg, while it was found to be $N_{2} =$ 0.086 for AD Leo \citep{Dal11}. According to these values, the flare frequencies of KIC 9641031 are definitely small. However, it is well known from \citet{Dal11} that the flare frequency dramatically changes from one season to the next for some stars, such as V1005 Ori, EV Lac, etc. Because of this, there could be some changes in the flare frequency and flare behaviour of KIC 9641031 in the next observing seasons.

On the other hand, this result obtained from the flare frequencies explained why any flare had not been detected from the system by any ground based telescope before the Kepler Mission. Although the flare frequency $N_{1}$ indicates that one flare occurs on the star per 2.402 hours, but $N_{2}$ frequency demonstrates that the energies of these flare are so small that it is very difficult to detect them with the ground based telescopes. Here it should be noted that, the parameters comparing from the literature were derived from the observation obtained in the standard Johnson U band. However, the observation data used in this study is different band.

The $Plateau$ value derived from the OPEA model of the flare equivalent duration distributions on the logarithmic scale versus the flare total durations was found to be 1.232$\pm$0.069 for 240 flares of KIC 9641031. According to \citet{Dal11}, this value is 3.014 for EV Lac ($B-V=1^{m}.554$) and 2.935 for EQ Peg ($B-V=1^{m}.574$), and also it is 2.637 for V1005 Ori ($B-V=1^{m}.307$). As it is seen that the maximum flare energy detected from KIC 9641031 is almost half of the maximum energy level obtained from UV Ceti type single flare stars. \citet{Dal11} found that the $Plateau$ value is always constant for a star, while it is changing from one star to the next depending on their $B-V$ color indexes. The authors defined the $Plateau$ value as the energy saturation level for the flare mechanism occurring on the target star.

The $half-life$ value was found to be 2291.7 s. This value is 10 times bigger than those found from UV Ceti type single flare stars. For instance, it was found to be 433.10 s for DO Cep ($B-V=1^{m}.604$), and 334.30 s for EQ Peg, while it is 226.30 s for V1005 Ori \citep{Dal11}. It means that in the case of the stars such as EQ Peg, V1005 Ori and DO Cep, the flares can reach the maximum energy level at their $Plateau$ value, when their total durations reach some 10 minutes, while it needs a few hours for KIC 9641031.

The similar extended durations are seen for the maximum flare rise and total times found from KIC 9641031, whereas these times obtained from UV Ceti type single stars are absolutely shorten than those seen in this system. For example, the maximum flare rise time was found to be 2062 s for V1005 Ori and 1967 s for CR Dra. However, it was found to be 5179 s for KIC 9641031. Similarly, the maximum flare total time was found to be 5236 s for V1005 Ori and 4955 s for CR Dra. In the case of KIC 9641031, it was obtained as 12770.62 s.

As a result, the flare activity level of KIC 9641031 is considerably lower than that seen in the others. However, this is in agreement with the results revealed by \citet{Dal11}. The authors demonstrated that the parameters derived from the OPEA model get values depending on the $B-V$ color index of the star. Therefore, according to the general trends found by \citet{Dal11}, the parameters of the KIC 9641031's flares are in agreement with the $B-V$ color index of the secondary component. This situation also indicates that the general trends found by \citet{Dal11} are valid around the spectral types of $B-V=0^{m}.74$. On the other hand, KIC 9641031 is an eclipsing binary system, so it is a double star. In this case, it is expected that the tidal interactions between the components make the magnetic activity level increase. However, it is not realized in the case of the flare activity patterns.

As seen from the literature, \citet{Bro10} given the age of the system between 3.05 Gy and 15.25 Gy. According to the relations among the age, rotational period and the magnetic activity level described by \citet{Sku72}, the given ages are too high that it should not be expected any high level magnetic activity on the components. Therefore, in the case of KIC 9641031, it is seen that being a component in a binary system does not affect the chromospheric activity as much as it is expected. Because, according to the semi-major axis ($a$) of the system, the components are too far away from each other to not affect the chromospheric activity.

On the contrary, it seems that KIC 9641031 has very high level spot activity unlike the flare activity. The variation due to rotational modulation is clearly revealed by the Kepler Mission with the highest quality sensitive observations \citep{Jen10a, Jen10b}.

The distribution of these spots on the surface was modelled by SPOTMODEL program \citep{Rib02, Rib03}. The analyses of the pre-whitened light curves indicate two cool spots on the one component for all 92 sub-data sets. The derived parameters of both spots, such as latitude ($b$), radius ($g$) and longitude ($l$) values, are listed in Table 3, while their variations versus time are shown in Figure 6. The latitudes of the spots are shown in the upper panel of the figure. As it can see, both spots are located between $+50^\circ$ and $+100^\circ$ in the latitude until HJD 24 56300. After this time, one of the spots is rapidly migrating to the latitude range from $-50^\circ$ to $-100^\circ$, while the other one is stable in that latitude.

In the model, locating of the stellar spots close to one of the poles solves out a problematic behaviour of the flare activity. If the phase distribution of the detected flares is examined, it is seen that there are the flare activity patterns in each phase interval. Although it is normally expected that large number of the flares should be seen in the phase interval, in which the observers directly see the spotted areas on the surface of the star, but this expectation is not working in this system. However, considering both the orbital inclination of 86$^\circ$.3 and the spotted area latitudes close to the pole, it is easy to understand that the active regions on the star are always in front of the observers. This situation explains why the flare patterns are seen in each phase interval.

As it is seen from the middle panel of the Figure 6, the longitudes of the spotted areas are overlapped and changed their sides between each other around HJD 24 56300, when the spots changed their locations in the latitude range. There are about 180$^\circ$ longitudinal differences between two spots in the beginning, while the longitudinal differences are decreasing set by set, and finally, two spots changed sides in the longitudinal plane around HJD 24 56300. This is very interesting phenomenon in the astrophysical sense, because the spot with bigger radius is migrating towards the earlier longitudes, while the second spot with smaller radius is migrating towards the later longitudes. The migrations of both spots get more distinctive according to each other after HJD 24 56300. According to \citet{Fek02} and \citet{Ber05}, the behaviour of the spot migration on a star is very important to understand both the rotation of the star and also the dynamo process working its inside.

The variations of the spot radii are seen in the bottom panel of the Figure 6. As seen from the figure, the radii of both spots sometimes increase and sometimes decrease. However, the remarkable point is that the radii of the spots are synchronously varying in opposite directions relative to each other. The radius of one spot is increasing, on the contrary, the radius of the other one is decreasing in the moment. This phenomenon seems to be a recurrent behaviour.

Like the synchronous longitudinal variations of the spots, the synchronous variations of the spot radii demonstrate that the assumption of "both spots are occurring on the same component" in the model is a right approach for this system. If the spots locate on the different components, it should not be expected any apparent synchronous variations like those.

Using the data obtained in the Kepler Mission, we demonstrated some clear variations with very short amplitudes in short time intervals. For example, the amplitude variations of the sinusoidal variation are lower than a few mmag, while it is clearly seen that this amplitude and light curve shapes are changing from one cycle to the next in short time intervals. However, in the case of observations made by the ground based telescopes, the variations of both the amplitude and the shape of the light curve due to the sinusoidal variations caused by rotational modulation can be perceived in a few months at least or generally from one season to the next \citep{Dal12}.

Considering the studies of \citet{Tra13} and \citet{Bal15}, it is expected that the chromospheric activity affects the orbital period of the KIC 9641031. Therefore, all minima times were computed from the short cadence data given in the database. After the linear correction applied to $(O-C)_{I}$ residuals, $(O-C)_{II}$ residuals were obtained. It is seen that the $(O-C)_{II}$ residuals exhibit a variation as expected. The stellar spot activity occurring on the active component leads the $(O-C)_{II}$ residuals of both the primary and secondary minima to vary synchronously, but in opposite directions, due to the effects presented by \citet{Tra13} and \citet{Bal15}. The dominant effect is seen in the $(O-C)_{II}$ residuals of the secondary minima. Moreover, as it is seen in the cases of the spot longitudes and radii, the variation character of the $(O-C)_{II}$ residuals is changed around HJD 24 56300. After this time, there is nearly no separation between $(O-C)_{II}$ residuals of the primary and secondary minima.

In this study, although the secondary component was assumed as chromospherically active component, it is possible that the primary component could be chromospherically active star, too. In this point, considering the parameter variations of the models, it is certain that there are two spotted areas and both of them are located on the same component.

Consequently, these results in the general respect reveal that KIC 9641031 is an active binary, but not active as much as the eclipsing binaries such as UV Ceti, BY Dra or RS CVn type variables, though one of the components exhibits both the flare and the cool spot activities. Because, both the flare energy level and the flare frequency are saliently lower than those obtained from UV Ceti type flare stars from the spectral type dMe \citep{Dal10, Dal11}. In addition, although both the minimum location and the shape of sinusoidal light variations due to the rotational modulation out-of-eclipses seem to be rapidly changing, but the amplitude of the variations is not large as much as that observed from other systems \citep{Dal12}. Nonetheless, KIC 9641031 has one chromospherically active component at least.

In the future, the spectral observations of the system should be done to certainly understand which component is an active star. In addition, the system should be observed photometrically with high time resolution to easily detect flares in order to check the flare frequencies, $N_{1}$ and $N_{2}$.

\section*{Acknowledgments} The authors thank Dr. O. \"{O}zdarcan for his help with the software and hardware assistance in the analyses. We also thank the referee for useful comments that have contributed to the improvement of the paper.

\clearpage

\begin{figure*}[h]
\begin{center}
\includegraphics[scale=0.75, angle=0]{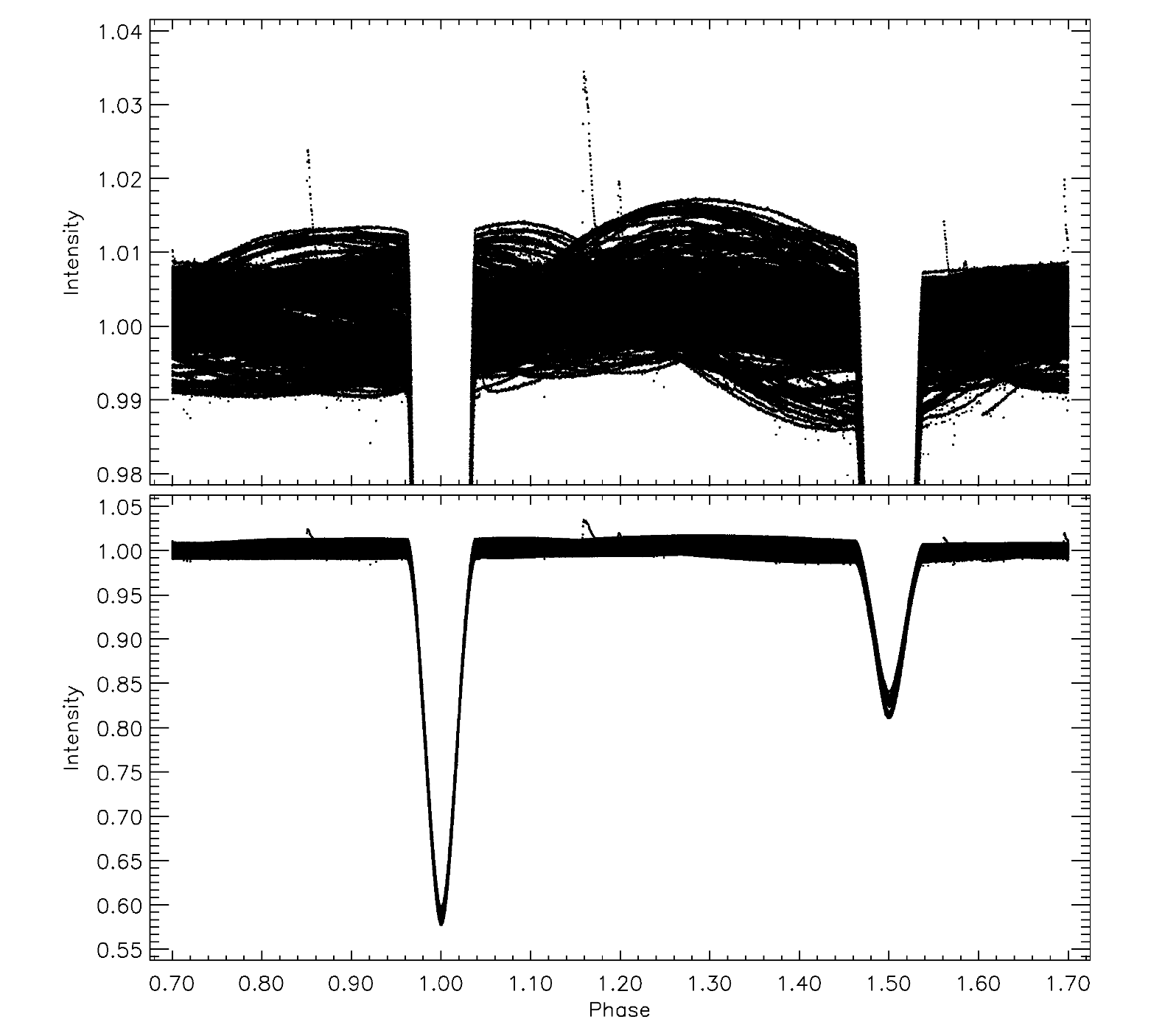}
\vspace{0.4 cm}
\caption{All the light curves obtained from the data given in the Kepler Mission Database are shown together with the error bars. The full of the light curves are shown in the bottom panel, while the maxima of the curves are shown in the upper panel to reveal the variations out of eclipses, such as sinusoidal variation due to the rotational modulation and sudden variations due to the flares.}
\label{Fig. 1.}
\end{center}
\end{figure*}

\begin{figure*}[h]
\begin{center}
\includegraphics[scale=0.75, angle=0]{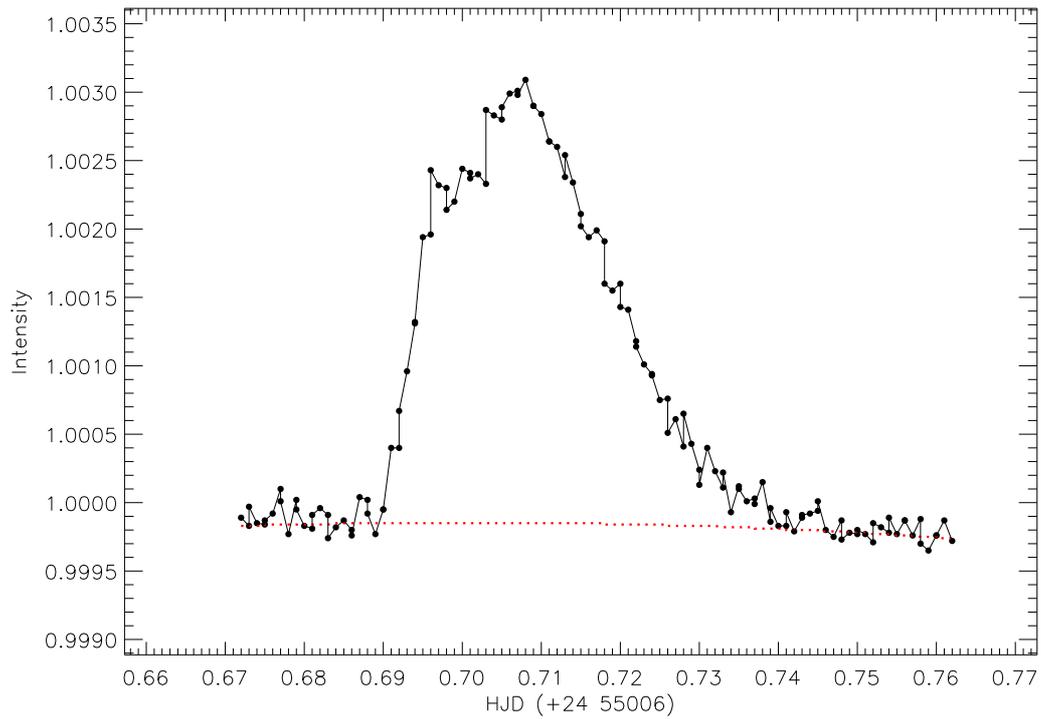}
\vspace{0.4 cm}
\caption{A slow flare example detected in the observations of the system. Filled circles show observations, while the dashed line represents the level of the quiescent state of the star for the observing night.}
\label{Fig. 2.}
\end{center}
\end{figure*}

\begin{figure*}[h]
\begin{center}
\includegraphics[scale=0.75, angle=0]{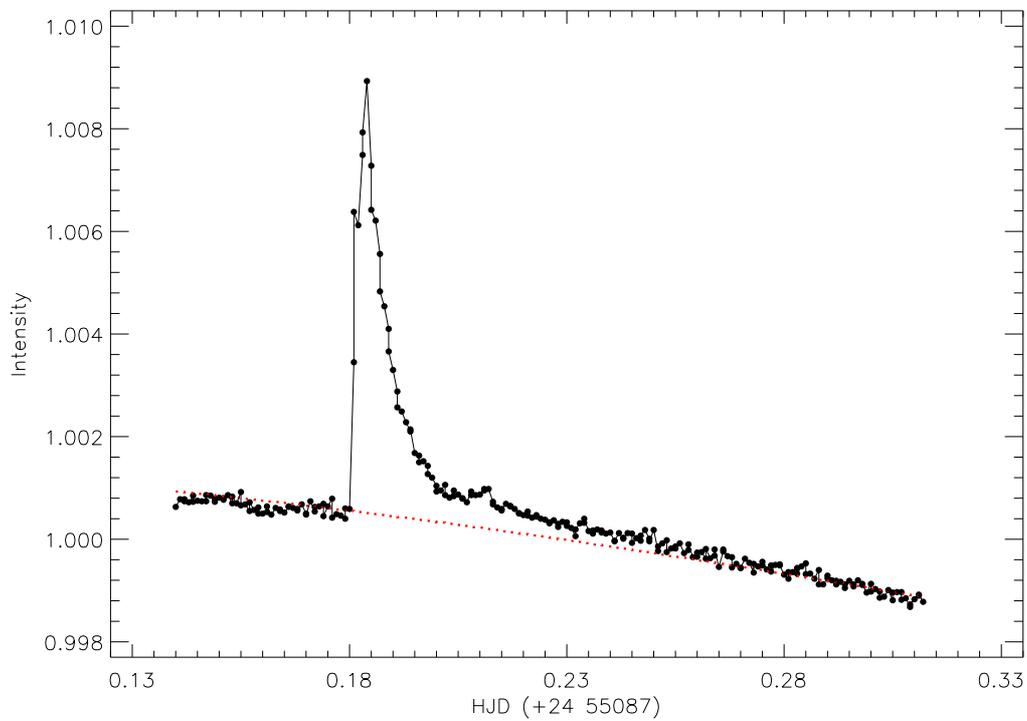}
\vspace{0.4 cm}
\caption{A fast flare example detected in the observations of the system. All the symbols are the same as in Figure 2.}
\label{Fig. 3.}
\end{center}
\end{figure*}

\begin{figure*}[h]
\begin{center}
\includegraphics[scale=0.75, angle=0]{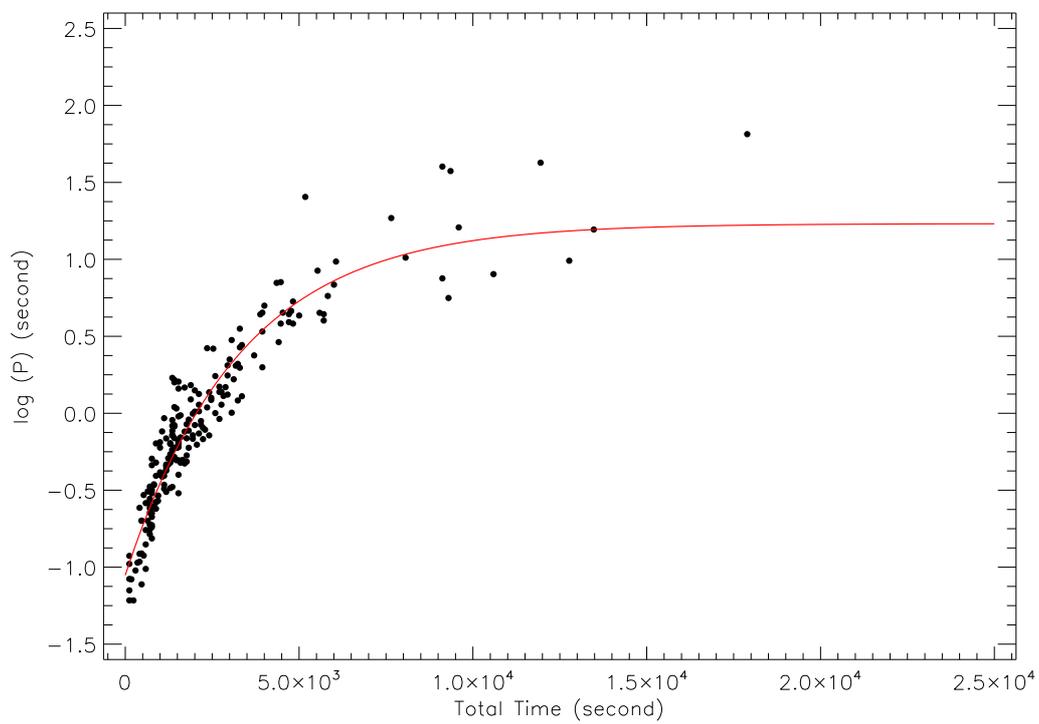}
\vspace{0.4 cm}
\caption{The distributions of flare equivalent durations on the logarithmic scale versus flare total durations for detected 240 flares and the OPEA model derived for this distribution. Filled circles show observed flares, while the line represents the OPEA model.}
\label{Fig. 4.}
\end{center}
\end{figure*}

\begin{figure*}[h]
\begin{center}
\includegraphics[scale=0.5, angle=0]{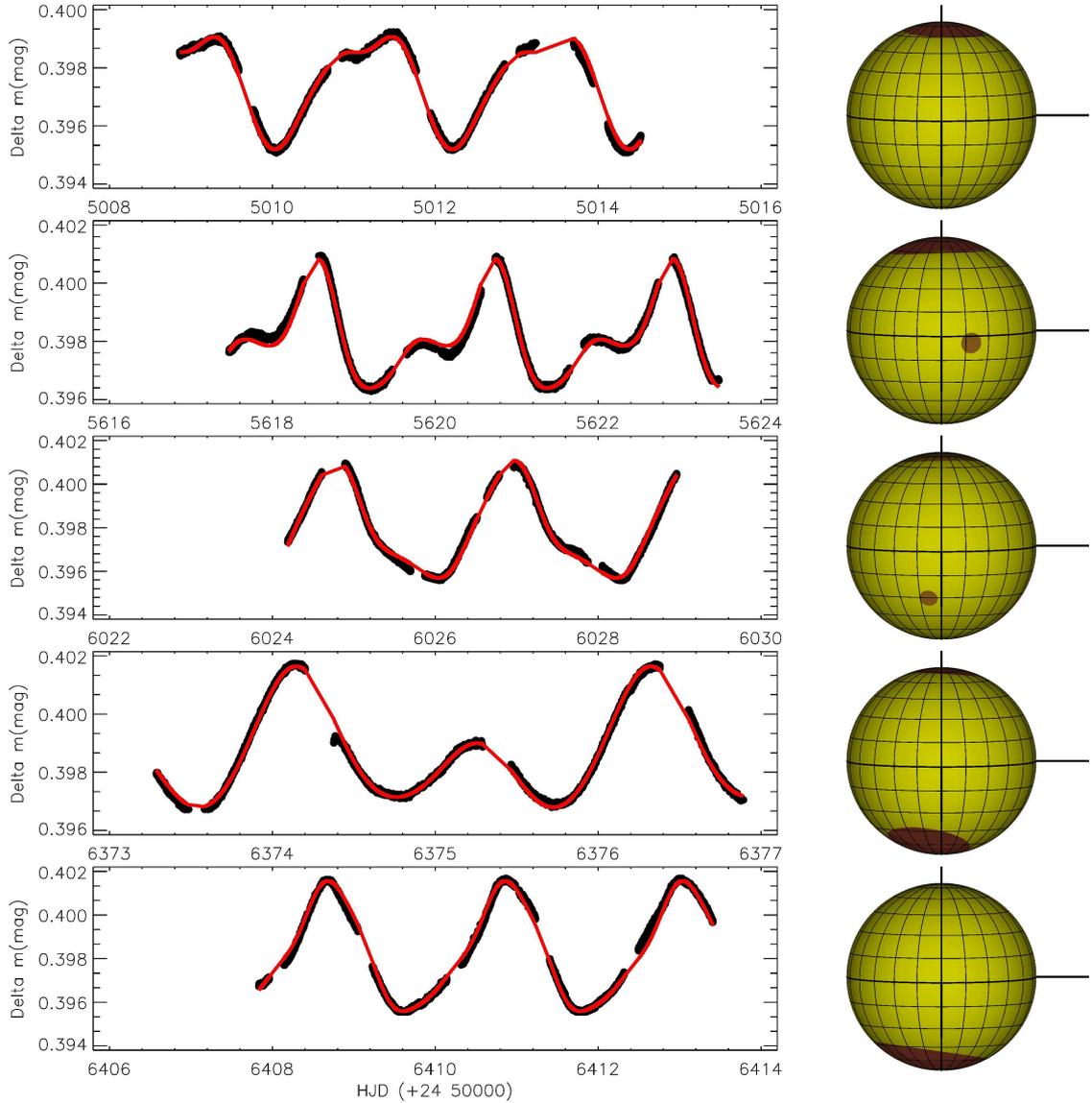}
\vspace{0.2 cm}
\caption{Some examples selected among all the models of rotation modulations due to cool spots. In the left panels, filled circles show observations arranged as the pre-whitened light curve, while the line represents the synthetic fits derived by the SPOTMODEL. In the right panels, the spot distributions on the active component surface derived by the SPOTMODEL are shown as the 3D form. In the figure, the pre-whitened light curve fit and its 3D model are shown side by side for the same sub-data set.}
\label{Fig. 5.}
\end{center}
\end{figure*}

\begin{figure*}[h]
\begin{center}
\includegraphics[scale=0.75, angle=0]{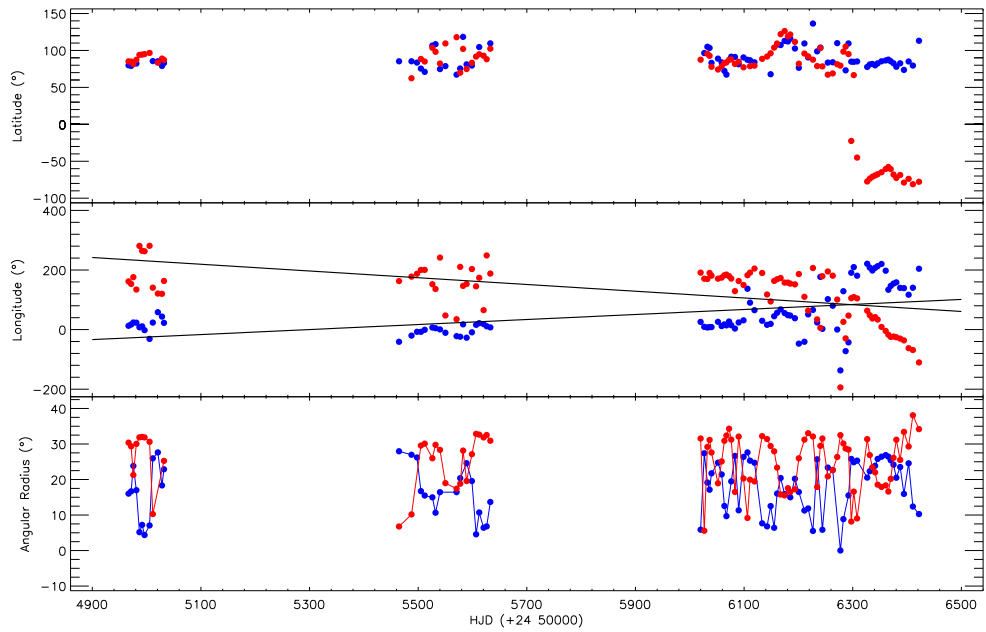}
\vspace{0.1 cm}
\caption{The variations of the parameters found by SPOTMODEL are shown. In the figure, filled red circles represent the Spot 1; filled blue circles represent the Spot 2. In the middle panel, two linear lines are shown just as the representative fits to the trends of spot longitudes. In the bottom panel, the filled circles are also consolidated with the thin lines to reveal the asynchronous trends between two spots.}
\label{Fig. 6.}
\end{center}
\end{figure*}

\begin{figure*}[h]
\begin{center}
\includegraphics[scale=0.75, angle=0]{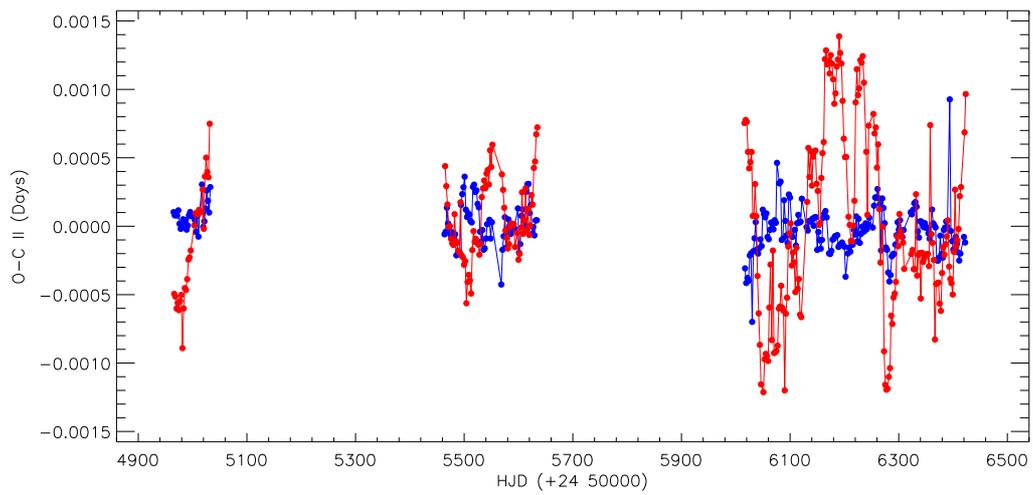}
\vspace{0.1 cm}
\caption{The variation of $(O-C)_{II}$ residuals obtained, after the linear correction on the $(O-C)_{I}$. In the figure, the filled blue circles represent the primary minima, while the filled red circles represent the secondary minima.}
\label{Fig. 7.}
\end{center}
\end{figure*}

\clearpage

\setcounter{table}{0}
\begin{table*}
\begin{center}
\caption{Calculated Parameters of Flares Detected in the Observations of KIC 9641031.}
\begin{tabular}{@{}ccrrr@{}}
\hline
\hline
Flare Time &	$P$ &	$T_{r}$ &	$T_{d}$ & Amplitude \\
(+24 00000)	& (s)	& (s)	& (s)	& 	 (Intensity)	\\
\hline									
54964.828101	&	0.463112	&	353.111616	&	882.763488	&	0.00085	\\
54966.495541	&	0.108480	&	58.857408	&	353.111616	&	0.00134	\\
54968.531480	&	0.326691	&	176.556672	&	1118.158560	&	0.00072	\\
54969.875376	&	0.182262	&	235.395936	&	529.658784	&	0.00081	\\
54970.999262	&	0.077346	&	235.405440	&	235.395936	&	0.00046	\\
54972.743670	&	0.530361	&	411.951744	&	941.610528	&	0.00073	\\
\hline
\end{tabular}
\end{center}
\end{table*}

\setcounter{table}{0}
\begin{table*}
\begin{center}
\caption{Continued From Previous Page.}
\begin{tabular}{@{}ccrrr@{}}
\hline
\hline
Flare Time &	$P$ &	$T_{r}$ &	$T_{d}$ & Amplitude \\
(+24 00000)	& (s)	& (s)	& (s)	& 	 (Intensity)	\\
\hline	
55502.426134	&	0.105038	&	58.835808	&	58.844448	&	0.00146	\\
55510.561799	&	0.243144	&	117.697536	&	294.223968	&	0.00140	\\
55511.945101	&	0.118927	&	58.853952	&	470.765952	&	0.00041	\\
55513.171751	&	0.225939	&	294.232608	&	411.922368	&	0.00092	\\
55526.031492	&	37.460548	&	1706.544288	&	7650.056448	&	0.01375	\\
55534.234607	&	0.386698	&	529.621632	&	529.622496	&	0.00080	\\
\hline
\end{tabular}
\end{center}
\end{table*}

\setcounter{table}{0}
\begin{table*}
\begin{center}
\caption{Continued From Previous Page.}
\begin{tabular}{@{}ccrrr@{}}
\hline
\hline
Flare Time &	$P$ &	$T_{r}$ &	$T_{d}$ & Amplitude \\
(+24 00000)	& (s)	& (s)	& (s)	& 	 (Intensity)	\\
\hline
56071.401857	&	0.179078	&	58.848768	&	58.858272	&	0.00269	\\
56075.048026	&	0.624893	&	294.254208	&	1765.526976	&	0.00082	\\
56075.977108	&	1.646454	&	235.396800	&	1177.026336	&	0.00269	\\
56079.256818	&	0.118439	&	58.849632	&	58.857408	&	0.00212	\\
56080.711061	&	2.236179	&	176.548032	&	2824.835040	&	0.00262	\\
56083.697196	&	9.794873	&	5178.865248	&	7591.754592	&	0.00238	\\
\hline
\end{tabular}
\end{center}
\end{table*}

\setcounter{table}{0}
\begin{table*}
\begin{center}
\caption{Continued From Previous Page.}
\begin{tabular}{@{}ccrrr@{}}
\hline
\hline
Flare Time &	$P$ &	$T_{r}$ &	$T_{d}$ & Amplitude \\
(+24 00000)	& (s)	& (s)	& (s)	& 	 (Intensity)	\\
\hline
56240.689558	&	1.698398	&	235.396800	&	1118.085984	&	0.00343	\\
56253.134493	&	0.579830	&	235.387296	&	1118.076480	&	0.00077	\\
56253.270712	&	3.909380	&	235.396800	&	4472.339616	&	0.00261	\\
56253.708655	&	0.577834	&	235.387296	&	1118.085120	&	0.00088	\\
56253.764505	&	0.333632	&	235.378656	&	470.775456	&	0.00092	\\
56259.634853	&	0.519913	&	588.456575	&	823.862016	&	0.00058	\\
\hline
\end{tabular}
\end{center}
\end{table*}

\setcounter{table}{1}
\begin{table*}
\begin{center}
\caption{The OPEA model parameters by using the least-squares method.}
\begin{tabular}{@{}lcc@{}}
\hline
\hline
Parameter	&	Values	&	95 Confidence Intervals	\\
\hline
$y_{0}$	&	-1.052$\pm$0.037	&	-1.1255 to -0.979	\\
$Plateau$	&	1.232$\pm$0.069	&	1.097 to 1.368	\\
$K$	&	0.0003$\pm$0.0001	&	0.0003 to 0.0003	\\
$Tau$	&	3306.2	&	2904.7 to 3836.5	\\
$Half-life$	&	2291.7	&	2013.4 to 2659.3	\\
$Span$	&	2.285$\pm$0.061	&	2.165 to 2.405	\\
		&     &			\\
\hline					
Goodness of Fit	&	Method	&	Values	\\
\hline					
$R^{2}$	&		&	0.9010	\\
$p-value$   	&	(D'Agostino-Pearson)	&	0.0010	\\
$p-value$   	&	(Shapiro-Wilk)	&	0.0005	\\
$p-value$   	&	(Kolmogorov-Smirnov)	&	0.0007	\\
\hline
\end{tabular}
\end{center}
\end{table*}

\setcounter{table}{2}
\begin{table*}
\begin{center}
\caption{The spot parameters obtained from the SPOTMODEL program.}
\begin{tabular}{@{}crrrrrr@{}}
\hline
\hline
HJD Average	&		$b_{Spot 1}$		&		$b_{Spot 2}$		&		$g_{Spot 1}$		&		$g_{Spot 2}$		&		$l_{Spot 1}$		&		$l_{Spot 2}$		\\
(+24 00000)	&		(degree)		&		(degree)		&		(degree)		&		(degree)		&		(degree)		&		(degree)		\\
\hline
54966.748617	&	79.990$\pm$1.342	&	85.249$\pm$0.731	&	16.027$\pm$1.850	&	30.379$\pm$0.971	&	12.693$\pm$1.401	&	161.691$\pm$1.850	\\
54971.548427	&	79.017$\pm$1.211	&	84.899$\pm$0.691	&	16.679$\pm$1.837	&	29.342$\pm$0.992	&	17.008$\pm$1.682	&	153.193$\pm$1.872	\\
54975.579032	&	82.600$\pm$0.455	&	81.536$\pm$0.925	&	23.829$\pm$0.674	&	21.287$\pm$1.410	&	23.755$\pm$1.304	&	176.001$\pm$1.927	\\
54981.174707	&	82.324$\pm$0.477	&	87.351$\pm$1.216	&	16.995$\pm$0.695	&	30.032$\pm$1.431	&	23.148$\pm$1.325	&	134.470$\pm$1.948	\\
54986.814506	&	83.559$\pm$0.498	&	94.028$\pm$0.012	&	5.178$\pm$0.135	&	31.869$\pm$0.016	&	7.753$\pm$0.163	&	280.838$\pm$0.447	\\
54991.711822	&	84.793$\pm$0.519	&	94.365$\pm$0.012	&	7.237$\pm$0.999	&	31.994$\pm$0.011	&	10.958$\pm$0.163	&	264.416$\pm$0.298	\\
\hline
\end{tabular}
\end{center}
\end{table*}

\setcounter{table}{2}
\begin{table*}
\begin{center}
\caption{Continued From Previous Page.}
\begin{tabular}{@{}crrrrrr@{}}
\hline
\hline
HJD Average	&		$b_{Spot 1}$		&		$b_{Spot 2}$		&		$g_{Spot 1}$		&		$g_{Spot 2}$		&		$l_{Spot 1}$		&		$l_{Spot 2}$		\\
(+24 00000)	&		(degree)		&		(degree)		&		(degree)		&		(degree)		&		(degree)		&		(degree)		\\
\hline
56211.056479	&	109.417$\pm$0.645	&	95.809$\pm$0.082	&	11.278$\pm$0.289	&	31.234$\pm$0.062	&	-40.907$\pm$0.174	&	110.014$\pm$0.396	\\
56217.956451	&	90.533$\pm$0.667	&	91.914$\pm$0.103	&	11.858$\pm$0.310	&	33.040$\pm$0.083	&	51.011$\pm$0.195	&	63.843$\pm$0.417	\\
56226.753235	&	136.389$\pm$0.688	&	87.493$\pm$0.039	&	5.517$\pm$0.328	&	32.116$\pm$0.032	&	66.005$\pm$0.302	&	206.760$\pm$0.955	\\
56234.493888	&	98.779$\pm$0.482	&	78.880$\pm$1.577	&	25.698$\pm$0.496	&	17.890$\pm$1.707	&	23.892$\pm$0.485	&	34.233$\pm$0.905	\\
56240.418359	&	103.924$\pm$1.839	&	103.262$\pm$0.524	&	26.932$\pm$1.149	&	29.476$\pm$0.381	&	177.001$\pm$0.509	&	5.803$\pm$0.294	\\
56244.159299	&	105.159$\pm$1.860	&	78.451$\pm$0.023	&	5.854$\pm$0.857	&	31.548$\pm$0.012	&	2.399$\pm$0.126	&	179.182$\pm$0.050	\\
\hline
\end{tabular}
\end{center}
\end{table*}

\setcounter{table}{3}
\begin{table*}
\begin{center}
\caption{Minima times and $(O-C)_{I}$ and $(O-C)_{II}$ residuals.}
\begin{tabular}{@{}cccccccccc@{}}
\hline
\hline
HJD	&	E	&	Type	&	$(O-C)_{I}$	&	$(O-C)_{II}$	&	HJD	&	E	&	Type	&	$(O-C)_{I}$	&	$(O-C)_{II}$	\\
(+24 00000)	&		&		&	(day)	&	(day)	&	(+24 00000)	&		&		&	(day)	&	(day)	\\
\hline
54965.02435	&	5.0	&	I	&	0.00087	&	0.00010	&	56094.39675	&	523.5	&	II	&	0.00026	&	-0.00052	\\
54966.11283	&	5.5	&	II	&	0.00027	&	-0.00049	&	56095.48623	&	524.0	&	I	&	0.00066	&	-0.00012	\\
54967.20248	&	6.0	&	I	&	0.00084	&	0.00008	&	56096.57528	&	524.5	&	II	&	0.00064	&	-0.00015	\\
54968.29097	&	6.5	&	II	&	0.00025	&	-0.00051	&	56097.66474	&	525.0	&	I	&	0.00102	&	0.00023	\\
54969.38064	&	7.0	&	I	&	0.00085	&	0.00009	&	56098.75353	&	525.5	&	II	&	0.00073	&	-0.00005	\\
54970.46903	&	7.5	&	II	&	0.00016	&	-0.00060	&	56099.84287	&	526.0	&	I	&	0.00100	&	0.00021	\\
\hline
\end{tabular}
\end{center}
\end{table*}

\setcounter{table}{3}
\begin{table*}
\begin{center}
\caption{Continued From Previous Page.}
\begin{tabular}{@{}cccccccccc@{}}
\hline
\hline
HJD	&	E	&	Type	&	$(O-C)_{I}$	&	$(O-C)_{II}$	&	HJD	&	E	&	Type	&	$(O-C)_{I}$	&	$(O-C)_{II}$	\\
(+24 00000)	&		&		&	(day)	&	(day)	&	(+24 00000)	&		&		&	(day)	&	(day)	\\
\hline
55465.99971	&	235.0	&	I	&	0.00073	&	-0.00004	&	56170.63419	&	558.5	&	II	&	0.00230	&	0.00121	\\
55467.08912	&	235.5	&	II	&	0.00107	&	0.00029	&	56171.72156	&	559.0	&	I	&	0.00059	&	-0.00020	\\
55468.17804	&	236.0	&	I	&	0.00091	&	0.00014	&	56172.81225	&	559.5	&	II	&	0.00220	&	0.00112	\\
55469.26714	&	236.5	&	II	&	0.00093	&	0.00016	&	56173.89971	&	560.0	&	I	&	0.00058	&	-0.00020	\\
55470.35608	&	237.0	&	I	&	0.00079	&	0.00002	&	56174.99024	&	560.5	&	II	&	0.00204	&	0.00125	\\
55471.44513	&	237.5	&	II	&	0.00077	&	0.00000	&	56176.07788	&	561.0	&	I	&	0.00061	&	-0.00018	\\
\hline
\end{tabular}
\end{center}
\end{table*}

\setcounter{table}{3}
\begin{table*}
\begin{center}
\caption{Continued From Previous Page.}
\begin{tabular}{@{}cccccccccc@{}}
\hline
\hline
HJD	&	E	&	Type	&	$(O-C)_{I}$	&	$(O-C)_{II}$	&	HJD	&	E	&	Type	&	$(O-C)_{I}$	&	$(O-C)_{II}$	\\
(+24 00000)	&		&		&	(day)	&	(day)	&	(+24 00000)	&		&		&	(day)	&	(day)	\\
\hline
55531.34426	&	265.0	&	I	&	0.00066	&	-0.00011	&	56237.06635	&	589.0	&	I	&	0.00075	&	-0.00004	\\
55532.43367	&	265.5	&	II	&	0.00099	&	0.00021	&	56239.24456	&	590.0	&	I	&	0.00081	&	0.00002	\\
55533.52250	&	266.0	&	I	&	0.00075	&	-0.00003	&	56240.33416	&	590.5	&	II	&	0.00133	&	0.00054	\\
55534.61189	&	266.5	&	II	&	0.00105	&	0.00028	&	56241.42267	&	591.0	&	I	&	0.00077	&	-0.00002	\\
55535.70051	&	267.0	&	I	&	0.00060	&	-0.00017	&	56242.51185	&	591.5	&	II	&	0.00087	&	0.00008	\\
55536.79010	&	267.5	&	II	&	0.00111	&	0.00033	&	56243.60086	&	592.0	&	I	&	0.00080	&	0.00001	\\
\hline
\end{tabular}
\end{center}
\end{table*}

\setcounter{table}{3}
\begin{table*}
\begin{center}
\caption{Continued From Previous Page.}
\begin{tabular}{@{}cccccccccc@{}}
\hline
\hline
HJD	&	E	&	Type	&	$(O-C)_{I}$	&	$(O-C)_{II}$	&	HJD	&	E	&	Type	&	$(O-C)_{I}$	&	$(O-C)_{II}$	\\
(+24 00000)	&		&		&	(day)	&	(day)	&	(+24 00000)	&		&		&	(day)	&	(day)	\\
\hline
55615.20329	&	303.5	&	II	&	0.00075	&	-0.00003	&	56310.03424	&	622.5	&	II	&	0.00048	&	-0.00031	\\
55616.29270	&	304.0	&	I	&	0.00108	&	0.00031	&	56322.01451	&	628.0	&	I	&	0.00089	&	0.00010	\\
55617.38160	&	304.5	&	II	&	0.00090	&	0.00013	&	56323.10328	&	628.5	&	II	&	0.00059	&	-0.00020	\\
55618.47086	&	305.0	&	I	&	0.00109	&	0.00031	&	56324.19265	&	629.0	&	I	&	0.00088	&	0.00009	\\
55619.55956	&	305.5	&	II	&	0.00071	&	-0.00006	&	56325.28146	&	629.5	&	II	&	0.00062	&	-0.00017	\\
55620.64880	&	306.0	&	I	&	0.00087	&	0.00010	&	56326.37083	&	630.0	&	I	&	0.00091	&	0.00012	\\
\hline
\end{tabular}
\end{center}
\end{table*}

\setcounter{table}{3}
\begin{table*}
\begin{center}
\caption{Continued From Previous Page.}
\begin{tabular}{@{}cccccccccc@{}}
\hline
\hline
HJD	&	E	&	Type	&	$(O-C)_{I}$	&	$(O-C)_{II}$	&	HJD	&	E	&	Type	&	$(O-C)_{I}$	&	$(O-C)_{II}$	\\
(+24 00000)	&		&		&	(day)	&	(day)	&	(+24 00000)	&		&		&	(day)	&	(day)	\\
\hline
56060.63579	&	508.0	&	I	&	0.00069	&	-0.00010	&	56386.26983	&	657.5	&	II	&	0.00066	&	-0.00013	\\
56061.72437	&	508.5	&	II	&	0.00019	&	-0.00059	&	56387.35907	&	658.0	&	I	&	0.00083	&	0.00004	\\
56062.81401	&	509.0	&	I	&	0.00076	&	-0.00003	&	56388.44804	&	658.5	&	II	&	0.00072	&	-0.00008	\\
56063.90283	&	509.5	&	II	&	0.00051	&	-0.00028	&	56389.53718	&	659.0	&	I	&	0.00079	&	0.00000	\\
56064.99222	&	510.0	&	I	&	0.00081	&	0.00003	&	56390.62631	&	659.5	&	II	&	0.00083	&	0.00004	\\
56066.08044	&	510.5	&	II	&	-0.00005	&	-0.00083	&	56392.80413	&	660.5	&	II	&	0.00050	&	-0.00030	\\
\hline
\end{tabular}
\end{center}
\end{table*}


\end{document}